\documentclass[12pt]{elsarticle}
\usepackage{amsfonts}
\usepackage{braket}
\usepackage{amssymb}
\usepackage{graphicx}
\usepackage[colorlinks=true,linkcolor=blue,urlcolor=black,citecolor=blue]{hyperref}
 
\journal{Physica B} 
 
\begin{document}

\begin{frontmatter}
\title{The distinctions between  the electrical conductivities under non-contact and contact current excitation in spin-split two-dimensional conductors}

\author{A.I.Kopeliovich}
\address{$^1$ B. Verkin Institute for Low Temperature Physics \& Engineering, NAS of Ukraine}

\author{P.V.Pyshkin}
\address{$^2$ Ikerbasque, Basque Foundation for Science, 48011, Bilbao, Spain}
\address{$^3$ Department of Theoretical Physics and History of Science, The Basque Country University (EHU/UPV), PO Box 644, 48080 Bilbao, Spain}

\author{A.N.Kalinenko}
\address{$^1$ B. Verkin Institute for Low Temperature Physics \& Engineering, NAS of Ukraine}

\author{A.V.Yanovsky}
\address{$^1$ B. Verkin Institute for Low Temperature Physics \& Engineering, NAS of Ukraine}

\begin{abstract}
It is shown that the normal electron-electron scattering is a source of electrical resistance on non-contact current excitation in two-dimensional spin-split electron systems. In contrast to the contact current injection, non-contact current excitation causes spatially inhomogeneous polarization in a two-dimensional conductor leading to new resistivity mechanisms. 
\end{abstract}

\begin{keyword}
Spin transport, Spintronics
\end{keyword}

\end{frontmatter}

It is well-known that in bulk conductors the ``normal'' electron-electron (e-e) collisions do not cause electrical resistance as they conserve the total momentum of the system of interacting particles. The picture however changes when spatial inhomogeneity occurs. For example, when electrical conductivity is influenced by the size effect the resistivity depends on the processes of momentum transfer to the rough boundaries of the sample. In this case, the electron-electron collisions may lengthen the path to the boundaries. As a result, the viscosity of the electron liquid increases and the electrical resistance decreases (the known Gurzhi-effect~\cite{ref_1, ref_2}).  The electrical conductivity of the electron system formed over the surface of superfluid helium (ESLHe) in the applied non-uniform magnetic field is another example when e-e scattering influences the electrical conductivity: the inhomogenity of the applied magnetic field leads to friction of the electron groups having different spins. As a result, the electron momentum is transferred to the magnetic field~\cite{ref_3, ref_4}.

The goal of this study is to demonstrate that the e-e collisions are an efficient resistivity mechanism when ac-current is induced to a spin-split degenerate low-dimensional conductor by a non-contact technique. This technique of current generation is practised in ESLHe experiments (see~\cite{ref_5}) and can be applied to other low-dimensional electron systems.

In contrast to the 3D case, the inhomogeneous polarization surface charge, induced with external electrodes in a low-dimensional electron system, is not separated spatially from the current carriers. It destroys the spatial homogeneity of a low-dimensional electron system and makes normal e-e scattering an effective mechanism of  relaxation of the electron momentum, which can be evident in the experimental data on electric conductivity. This is of particular importance since e-e collisions dominate over other scattering processes in a number of low-dimensional electron systems~\cite{ref_5, ref_6}. Spin splitting can be reached by the exchange electron-electron interaction or by an applied uniform magnetic field. To avoid orbital effects we assume that some uniform magnetic field is applied parallel to the conducting plane.

Note that many low-dimensional conducting systems are ``good'' conductors~\cite{ref_3, ref_4, ref_7}, much like 3D metals~\cite{ref_6}, because their sizes along the conducting direction far exceed the corresponding characteristic screening radii in these systems and the typical experimental frequencies of applied fields are considerably lower than the characteristic plasma ones. The screening radius in ESLHe,~$r_0$ , is about~$T/e^2\rho$, where~$T$ is the temperature,~$e$ is the electron charge and~$\rho$ is the electron density~\cite{ref_9}.  The typical electron densities and temperatures of a ESLHe are as follows:  $\rho\approx10^8$~cm$^{-2}$ and $T\approx 1$~K. This yields $r_0\approx10^{-5}$~сm. When the frequency of the applied field is~$\omega << (4\pi^2e^2\rho/mL_c)^{1/2}$ and the conductor length $L_c>>r_0$, it is a ``good'' conductor; for example,  for~$L_c\approx 1$~cm we have the condition~$\omega << 10^8$~s$^{-1}$. In the case of degenerate two-dimensional electron gas, when~$\rho\approx10^{11}$~cm$^{-2}$ and~$L_c\approx10^{-3}$~cm, the above condition is~$\omega<< 10^{11}$~s$^{-1}$. 

In contrast to 3D metals, the polarization surface charge, $\rho_p$, in a low-dimensional conducting system is induced inside conducting region. It is the charge that ensures a constant electrical potential. Hence, the condition of electric neutrality yields

\begin{equation}\label{eq1}
\Delta\rho_\sigma = \delta\rho_\sigma + \frac{\Pi_\sigma}{\Pi}\rho_p, \;\; \sum_\sigma\delta\rho_\sigma = 0.
\end{equation}

Here $\Delta\rho_\sigma$ is the equilibrium addition to the spin-dependent density $\rho_\sigma$ (the equilibrium density  in the absence of the applied alternating electrical field), $\Pi_\sigma\rho_p/\Pi$~is the part of~$\Delta\rho_\sigma$ which is the equilibrium at a given instant in respect of the alternating electric field, $\sigma$~is the spin index, corresponding to different spin components, i.e. spin-up and spin-down electrons. $\Pi_\sigma$~is the spin-dependent density of states, $\Pi=\sum_{\sigma}\Pi_\sigma$;  $\delta\rho_\sigma$~is the non-equilibrium part of~$\Delta\rho_\sigma$.

Let us discuss the case when the spin-relaxation processes are not essential. Then the electrical conductivity is determined by the processes  of spin diffusion and the pressure of non-equilibrium spin components. The spin-dependent transport and spin diffusion are analyzed in terms of the quasi-classical two-liquid hydrodynamic approach~\cite{ref_2, ref_9}. The hydrodynamic approach can be applied assuming that the electron mean free path is much shorter than the  inhomogeneity dimension~$L$. In the linear ac field approximation the following hydrodynamic equations can be written for the spin-dependent densities and their currents~$j_\sigma$ (see Refs.~\cite{ref_3, ref_4, ref_11})
\begin{eqnarray}
i\omega\Delta\rho_\sigma + j_\sigma' = - \nu_s\Pi^*(\mu_\sigma - \mu_{-\sigma}) \label{eq2} \\
(i\omega + \nu)mu_\sigma + (\delta\mu_\sigma + e\delta\varphi)' = - m\nu_{ee}\frac{\rho_{-\sigma}}{\rho_\sigma}(u_\sigma - u_{-\sigma}) \label{eq3} \\
\Pi_\sigma\delta\mu_\sigma = \delta\rho_\sigma, \label{eq4} \\
\Pi^{*-1}=\sum_{\sigma}\Pi_\sigma^{-1} \nonumber 
\end{eqnarray}
For simplicity, we consider the case when all the functions depend on one coordinate only – precisely the $x$-coordinate. Then the prime marks differentiation with respect to~$x$. Here $\omega$ is the frequency of the applied electrical field; $\rho=\sum_{\sigma}\rho_\sigma$~is the initial total density, $\delta\varphi$~is the non-equilibrium addition to the potential of the ac electrical field due to the electrical current flow; $\delta\mu_\sigma$~is the non-equilibrium addition to the electrochemical potential for the corresponding spin component and $u_\sigma = j_\sigma/\rho_\sigma$~is the corresponding drift velocity; $\nu$~is the frequency of electron collisions without conserving the electron momentum, $\nu_{ee}$~is the frequency of electron-electron collisions, $\nu_s$~is the frequency of the spin-flip scattering. Note that the equilibrium part of the alternating field, $\varphi_e'$, is absent in Eq.~(\ref{eq3}) since it is balanced by the equilibrium pressure of the spin-dependent components of the polarization charge, i.e.~$e\varphi_e + (\Delta\rho_\sigma - \delta\rho_\sigma)/\Pi_\sigma =0$. In contrast to the standard hydrodynamic equations for the mixture of liquids, we speculate that the velocities of different spin components, $u_\sigma$, are different in magnitude.

Assuming a given dependence of the polarization surface charge on the $x$-coordinate, we can readily obtain the solution of Eqs.~(\ref{eq1})-(\ref{eq4}). Equation for the velocity difference, $\Delta(u) = u_\uparrow - u_\downarrow$  (the ``up'' and ``down'' arrows correspond to one of spin components), can be written as
\begin{equation}\label{eq5}
m(i\omega + \nu_s)[i\omega + \nu + \nu_{ee}]\Delta(u) = \frac{\rho_\uparrow\rho_\downarrow}{\rho}\Pi^*\Delta''(u) + \frac{j_p''}{\rho}\Delta(\rho_\sigma/\Pi_\sigma)
\end{equation}
Here $j_p$ is the polarization current. The following designation is introduced: $\Delta(a)=a_\uparrow - a_\downarrow$. According to Eqs.(\ref{eq1})-(\ref{eq2}), $j_p' = -i\omega\rho_p $.

Pursuing the goal of revealing new effect, we reduced the discussion to two limiting cases of the exact solution of Eq.~(\ref{eq5}) with the boundary condition which corresponds to the zero current through the boundaries of the conducting channel.

The first asymptotic solution is for the case of low frequencies of the ac field $\omega<<D/L^2$, where $D = (\rho_\uparrow\rho_\downarrow/\rho)\Pi^*/( m[i\omega + \nu + \nu_{ee}] )$ is the coefficient of spin diffusion. In this case the spin diffusion processes have enough time to attain the equilibrium state inside each spin subsystem. We can than write 
\begin{equation}\label{eq6}
\Delta(u) = -\frac{j_p\Pi^*}{\rho_\uparrow\rho_\downarrow}\Delta(\rho_\sigma/\Pi_\sigma)
\end{equation}
In the opposite case $\omega>>D/L^2 $, Eq. (\ref{eq5}) yields
\begin{equation}\label{eq7}
\Delta(u) = \frac{j_p''\Delta(\rho_\sigma/\Pi_\sigma)}{m\rho(i\omega+\nu_s)(i\omega+\nu+\nu_{ee})}
\end{equation}
An electrical field $-\delta\varphi'$ inducing the electrical current~$ej_p$ can be written in terms of the difference between the drift velocities~$\Delta(u)$ as
\begin{equation}\label{eq8}
\delta\varphi' = -\frac{m}{e\Pi\rho}\left[ \vphantom{1^1} \Delta(\Pi_\sigma\rho_{-\sigma})(i\omega + \nu +\nu_{ee})\Delta(u) + (i\omega + \nu)\Pi j_\sigma  \right]
\end{equation}

\begin{figure}[htbp]
	\begin{center}
		\includegraphics[width=8.5cm]{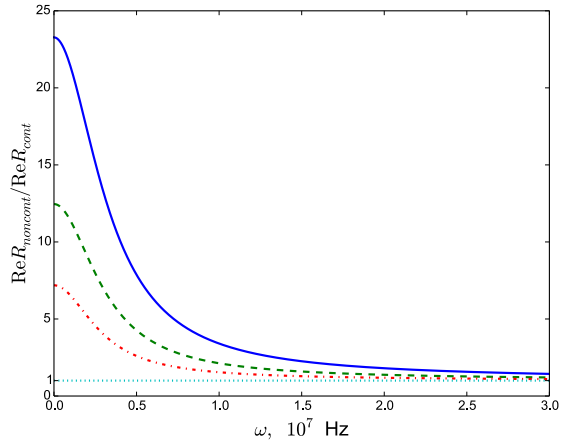}
	\end{center}
	\caption{(Color online) The frequency dependence of the ratio between the real parts of resistances under non-contact and contact current excitation. $\nu_{ee}=10^{11}$~sec$^{-1}$, $\nu=10^{10}$~sec$^{-1}$, $\nu_{s}=0$, $L=10^{-2}$~cm, $\rho_\uparrow=10^{11}$~cm$^{-2}$; blue curve: $\rho_\uparrow/\rho_\downarrow=1/10$, green curve: $\rho_\uparrow/\rho_\downarrow=1/6$; red curve: $\rho_\uparrow/\rho_\downarrow=1/4$.} \label{fig1}
\end{figure}
The first term in the square brackets in the right-hand side of Eq.~(\ref{eq8}) corresponds to the pressure of non-equilibrium spin components and the electron-electron scattering term describes friction of the spin components. The second term is the electrical resistivity of the conductor on direct contact connection to the source of the potential difference (a spatially uniform case). According to Eqs.~(\ref{eq6}),~(\ref{eq7}), the drift velocities of spin-up and spin-down components are different when~$\Delta(\rho_\sigma/\Pi_\sigma)\neq 0$. The difference is large in comparison with the mean velocity~$j_p/\rho $  when the spin components have enough time to adjust themselves to the applied ac field but the difference decreases when the frequency of the field increases and the adjusted state is disturbed.

The frequency dependence of the ratio between the real parts of resistances~$\mathrm{Re}(R_{noncont})/\mathrm{Re}(R_{cont})$ is demonstrated in Fig. 1. It is seen that the two methods of current excitation differ essentially when the electron-electron scattering predominates over the scattering processes~$\nu_{ee}>>\nu$ and the spin polarization is high enough. However, in the absence of the electron-electron scattering, the frequency dependences of~$R_{noncont}$ and~$R_{cont}$ are different too, see~Eq.~(\ref{eq8}). 

As follows from Eqs.~(\ref{eq6}), (\ref{eq8}), the electrical resistance increases when one of the spin densities vanishes (we take into account that the density of states of the degenerated 2D system does not tend to zero in this case). Reason is that the processes of spin diffusion slow down. Indeed, the electrons of the dominant spin polarization group have to ``wait'' for when the electrons of the minor group start to move because the opposite diffusion flows of  both groups must be equivalent. Otherwise the electric neutrality will be disturbed. The mentioned increase in the resistance at a fixed frequency of the applied field~$\omega$ is limited by the disturbance of the quasiequilibrium, i.e. the condition~$\omega<<D/L^2$.

When spin splitting is caused by the applied magnetic field, we have~$\Delta_s = 2\mu_B H$, where $\mu_B$~is the Bohr magneton. For the typical electron density of ESLHe in the liquid phase~$10^8$ cm$^{−2}$ we obtain~$\rho\hbar^2/Tm\approx 10^{-3}$. It is hardly probable that the discussed difference between the contact and noncontact injection can be observed in this case. The electron crystal on a surface of superfluid helium at temperatures below~1~K calls for a special discussion. In the latter case the electron-electron interaction is essential too but it cannot be considered in terms of electron-electron collisions (see~Ref.~\cite{ref_13}).

When the temperature is much lower than the Fermi energy~$\varepsilon_F$ (typically in experiments with two-dimensional electron gas in semiconductor structures) and the spin splitting of the electron energy, $\Delta_s$, is about~$\varepsilon_F$, $\Delta_s\approx\varepsilon_F$ , we have~$\Delta(\rho_\sigma/\Pi_\sigma)\approx\rho/\Pi$.

Note that in the 3D case, this type of effects on noncontact current excitation can be observed near the surface at the depth about the mean free path of spin relaxation, i.e.~$ \sqrt{D/(\nu_s+\omega)}$. This is because the equilibrium of the spin components of the surface and volume charges is disturbed during repolarization.

Note that the same type of effects can be expected in one-dimensional conductors. It is well-known~\cite{ref_14} that the electron-electron interaction is essential here but it must be described beyond the Fermi-liquid theory.

Finally we should note that the conclusion on the difference between the electric conductivities on noncontact and contact current excitation is based on ``good'' conductivity of the media and the availability of two groups of weakly coupled current carriers which attain equilibrium much slower in comparison with the spin diffusion. Spin splitting provides a favorable conditions for this because the spin flip processes are relativistic and rather slow.

In summary, it is shown that at non-contact current excitation the electrical resistance of a two-dimensional spin-split conductor exhibits a normal electron-electron collisions and spin-flip scattering.

\section*{ACKNOWLEDGMENTS}

This research was made possible partly by the NASU Nanoprogram Grant No.4/15-N.

\bibliographystyle{elsarticle-num}


\end{document}